\documentclass[letter,twocolumn]{jpsj3}
\usepackage{graphicx}
\usepackage[dvips]{color}
\usepackage{braket}
\usepackage{color} 
\usepackage{ulem}

\title{Weak Ferromagnetism in NiS$_2$ under Nanocrystallization}

\author{Hayato Miyazaki$^1$, Tomohiko Yoshinaga$^1$, Ojiro Miyazaki$^1$,  Akira Matsuo$^2$, Koichi Kindo$^2$, Hirofumi Ishii$^3$, Tatsuya Kawae$^4$, Tetsuya Kida$^5$, Masashi Nantoh$^{6, 7}$, and Yoichi Ishiwata$^1$\thanks{ishiwata@cc.saga-u.ac.jp}}
\inst{$^1$Department of Physics, Saga University, Saga 840-8502, Japan \\
$^2$Institute for Solid State Physics, University of Tokyo, Kashiwa, Chiba 277-8581, Japan \\
$^3$National Synchrotron Radiation Research Center, Hsinchu 30076, Taiwan \\
$^4$Department of Applied Quantum Physics, Kyushu University, Fukuoka 819-0395, Japan \\
$^5$Department of Applied Chemistry and Biochemistry, Kumamoto University, Kumamoto 860-8555, Japan \\
$^6$Pioneering Research Institute, The Institute of Physical and Chemical Research (RIKEN), Wako, 351-0198, Japan \\
$^7$RIKEN Nishina Center for Accelerator-Based Science (RNC), Wako, 351-0198, Japan} 

\abst{Structurally well-ordered NiS$_2$ nanocrystals with an average diameter of \(27.0 \pm 6.5\) nm retain the bulk-like two-step antiferromagnetic transitions, as shown by magnetization and heat-capacity measurements. Below the lower transition, the nanocrystals exhibit a hysteretic ferromagnetic response with large coercivity, exchange bias, and a vertical loop shift after field cooling, whereas the \(M\)-\(H\) response just above the transition is nearly linear. Taken together, these features are best explained by uncompensated surface moments generated where the low-temperature antiferromagnetic order terminates at the nanocrystal surface. The absence of a clear additional bulk-like weak-ferromagnetic component constrains homogeneous-canting models and indirectly favors a domain-wall scenario for the weak ferromagnetism of bulk NiS$_2$.}

\begin{document}
\maketitle

NiS$_2$ is a pyrite-type Mott insulator in which Ni ions form an fcc sublattice, providing a frustrated network for antiferromagnetic interactions. Bulk NiS$_2$ undergoes two successive magnetic transitions. Below \(T_{\rm N1}\simeq39\) K, a type-I antiferromagnetic order with propagation vector \(\boldsymbol{k}=(000)\) develops, forming a noncoplanar four-sublattice spin structure, and continues to grow on cooling. Below \(T_{\rm N2}\simeq30\) K, a type-II antiferromagnetic component with \(\boldsymbol{k}=(1/2,1/2,1/2)\) appears in addition to the type-I order, producing a complex low-temperature magnetic state accompanied by weak ferromagnetism \cite{Miyadai,Thio, Matsuura,Yano}. Although this weak-ferromagnetic state has been known for decades, its microscopic origin remains unresolved. Because pyrite NiS$_2$ is centrosymmetric, the usual Dzyaloshinskii--Moriya canting mechanism is not an obvious explanation. Yoshimori and Fukuda proposed a NiS$_2$-specific canted weak-ferromagnetism model based on a free-energy term allowed in the coexistence of type-I and type-II antiferromagnetic components \cite{YoshimoriFukuda}. A small structural distortion accompanying the transition at \(T_{\rm N2}\) has also been reported and proposed to induce a slight canting of the antiferromagnetic moments, thereby producing the weak-ferromagnetic moment \cite{Kikuchi,Thio,Higo}.

Magnetic domain walls provide another possible source of ferromagnetic moments in NiS$_2$. In all-in--all-out-type antiferromagnets, domain walls between time-reversal-related magnetic domains can act as internal interfaces carrying in-gap states and net magnetic moments \cite{Yamaji}. Robust ferromagnetic moments at antiferromagnetic domain walls have also been demonstrated in Cd$_2$Os$_2$O$_7$ \cite{Hirose}. For NiS$_2$, Higo and Nakatsuji observed a small, nearly isotropic magnetic irreversibility in the intermediate noncoplanar antiferromagnetic (NAF) phase between \(T_{\rm N2}\) and \(T_{\rm N1}\), and attributed this component to ferromagnetic moments at antiferromagnetic domain walls \cite{Higo}. They further showed that field cooling through the low-temperature weak-ferromagnetic phase controls the domain configuration retained in the NAF phase upon subsequent warming, indicating a close correspondence between the domains in the two phases. Although their study did not assign the main weak-ferromagnetic component below \(T_{\rm N2}\) to domain-wall moments, this correspondence motivates us to examine whether magnetic domain walls may also be involved in the microscopic origin of the low-temperature weak ferromagnetism.

NiS$_2$ nanocrystals (NCs) provide a useful way to test these two pictures. 
Although finite-size effects can shift or round magnetic transitions, antiferromagnetic order generally survives in particles with diameters of several tens of nanometers \cite{Fisher,He,Thota,Lang}. If the weak ferromagnetism of bulk NiS$_2$ originates from antiferromagnetic domain walls, this contribution is expected to be strongly suppressed in NCs, because creating internal magnetic domains generally requires a domain-wall energy cost, and sufficiently small particles tend to avoid multidomain states \cite{Kittel}. Indeed, all-in--all-out antiferromagnetic domains and domain walls have been directly visualized as extended real-space structures in Cd$_2$Os$_2$O$_7$ \cite{Tardif} and Nd$_2$Ir$_2$O$_7$ \cite{Ma}. In contrast, a spatially homogeneous canted moment should survive as a volume contribution as long as the underlying antiferromagnetic order is retained. Thus, the survival or disappearance of the weak-ferromagnetic response in NiS$_2$ NCs provides a test of whether the weak ferromagnetism is a homogeneous property of the low-temperature phase or is tied to extended magnetic boundaries.

Here, we synthesize structurally well-ordered NiS$_2$ NCs and investigate their structural, thermodynamic, and magnetic properties. The NCs retain the bulk-like two-step antiferromagnetic transitions. A hysteretic ferromagnetic response appears below \(T_{\rm N2}\), but it is accompanied by much larger coercivity than in bulk NiS$_2$ and by field-cooling-induced horizontal and vertical loop shifts. This contrast allows us to discuss how nanocrystallization constrains the microscopic origin of weak ferromagnetism in NiS$_2$.

\begin{figure}
\begin{center}
\includegraphics[scale=0.45]{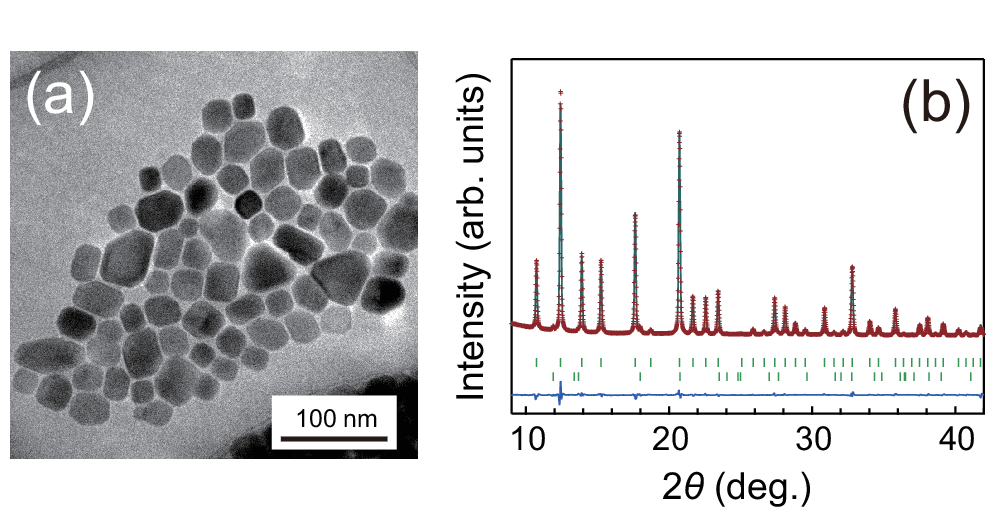}
\end{center}
\caption{(Color online) Structural characterization of NiS$_2$ NCs. (a) TEM image. (b) Synchrotron powder XRD pattern and Rietveld refinement profile.
}
\label{f1}
\end{figure}

NiS$_2$ NCs were synthesized by solvothermal treatment of Ni(II) acetylacetonate (0.252 g) and sulfur (0.320 g) in oleylamine (10 mL) at \(260^\circ\)C for 5 h. A representative TEM image is shown in Fig.~1(a). The average particle diameter evaluated from TEM images is \(27.0 \pm 6.5\) nm.

Synchrotron  x-ray diffraction (XRD) measurements were performed at BL12B2 of SPring-8 using a wavelength of 0.6199 \AA, and the data were analyzed by Rietveld refinement using RIETAN-FP \cite{Izumi}. Figure~1(b) shows the synchrotron powder XRD pattern and Rietveld refinement profile. The refinement was performed using a two-phase model consisting of cubic pyrite-type NiS$_2$ and hexagonal NiAs-type NiS, yielding a minor NiS phase fraction of 2.4\%. The lattice constant and S positional parameter of the NiS$_2$ phase were refined to be \(a=5.68667(5)\) \AA\ and \(x_{\rm S}=0.39447(5)\), respectively, close to the bulk values~\cite{Nowack}.

DC magnetization measurements were performed using a superconducting quantum interference device magnetometer (MPMS, Quantum Design). Temperature-dependent field-normalized magnetization \(M/H\) was measured under zero-field-cooled (ZFC) and field-cooled (FC) protocols. Magnetic-field-dependent magnetization curves were also measured after ZFC and FC procedures using the same system. Specific-heat measurements were performed using a physical property measurement system (PPMS, Quantum Design) by the relaxation method.

\begin{figure}
\begin{center}
\includegraphics[scale=0.7]{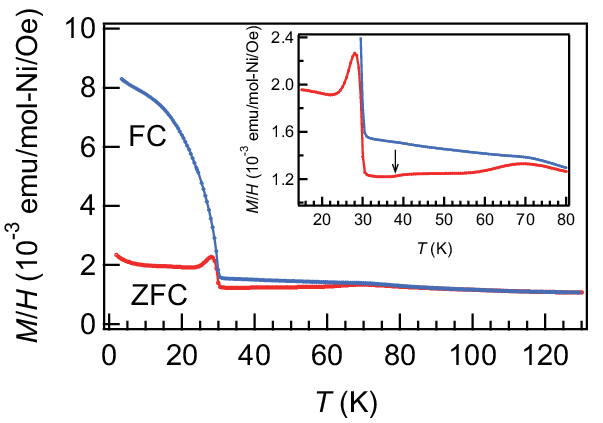}
\end{center}
\caption{(Color online) ZFC and FC \(M/H\) curves of NiS$_2$ NCs measured at \(H=1000\) Oe. The inset shows an enlarged view around the two antiferromagnetic anomalies; the arrow marks the \(T_{\rm N1}\)-related anomaly.}
\label{f2}
\end{figure}

Figure~2 shows the temperature dependence of \(M/H\) measured at \(H=1000\) Oe under ZFC and FC conditions. The ZFC and FC curves start to separate around 70 K, and a marked increase in the FC magnetization appears below approximately 30 K. The ZFC curve shows a peak slightly below 30 K, while both the ZFC and FC curves exhibit a small upturn below approximately 10 K. A small anomaly associated with the upper transition of bulk NiS$_2$, \(T_{\rm N1}\simeq39\) K, is difficult to discern in the raw \(M/H\) curves, as is also the case for bulk NiS$_2$~\cite{Higo}. Nevertheless, the enlarged view in the inset reveals a small increase in the ZFC curve just below 40 K. The more pronounced change near 30 K is associated with the lower transition, \(T_{\rm N2}\), where the type-II antiferromagnetic component appears on top of the type-I order and weak ferromagnetism emerges.

To quantify the anomaly temperatures, we calculated \(d[(M/H)T]/dT\) from the ZFC data using Fisher's relation \(d(\chi T)/dT\)~\cite{Fisher1962}. Empirical Lorentzian fits to the derivative anomalies associated with the two bulk-like transitions give estimated anomaly temperatures of \(T_{\rm N1}=39.4\) K and \(T_{\rm N2}=29.8\) K. These values show that the two-step antiferromagnetic transitions of bulk NiS$_2$ are retained in the NCs.

The present data do not allow us to determine whether the magnetic irreversibility attributed by Higo and Nakatsuji to domain-wall moments in the intermediate NAF phase is present in the NCs, because the FC--ZFC difference is superimposed on a bifurcation already present above \(T_{\rm N1}\). Separately, Thio et al. showed that surface contributions enhance the susceptibility of NiS$_2$ above \(T_{\rm N1}\) \cite{Thio}. In fact, at 50 K, the ZFC value of \(M/H\) in the present NCs is approximately \(1.25\times10^{-3}\) emu/mol-Ni/Oe, which is higher than both the value of \(7.2\times10^{-4}\) emu/mol-Ni/Oe reported for a high-quality flux-grown single crystal and the previously reported bulk values of approximately \(7.6\)--\(9.2\times10^{-4}\) emu/mol-Ni/Oe \cite{Higo}.

\begin{figure}
\begin{center}
\includegraphics[scale=0.6]{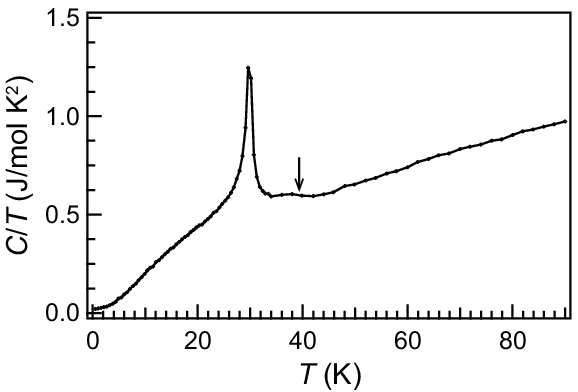}
\end{center}
\caption{Temperature dependence of \(C/T\) for NiS$_2$ NCs. The arrow marks the \(T_{\rm N1}\)-related anomaly.}
\label{f3}
\end{figure}

Figure~3 shows the temperature dependence of \(C/T\) for NiS$_2$ NCs. Two heat-capacity anomalies are observed at temperatures consistent with those estimated from \(d[(M/H)T]/dT\). The arrow marks the anomaly associated with \(T_{\rm N1}=39.4\) K, while a more pronounced anomaly is observed at \(T_{\rm N2}=29.8\) K. These results confirm that the two-step antiferromagnetic transitions of bulk NiS$_2$ are retained in the NCs.

The other magnetic features observed in Fig.~2, including the ZFC--FC bifurcation around 70 K, the ZFC peak just below \(T_{\rm N2}\), and the low-temperature upturn below approximately 10 K, do not show corresponding sharp anomalies in \(C/T\). Thus, these features are unlikely to represent additional bulk thermodynamic phase transitions. The high-temperature bifurcation and the ZFC peak below \(T_{\rm N2}\) are discussed below in connection with magnetic hysteresis, whereas the low-temperature upturn is likely due to a localized-spin contribution.

\begin{figure}
\begin{center}
\includegraphics[scale=0.4]{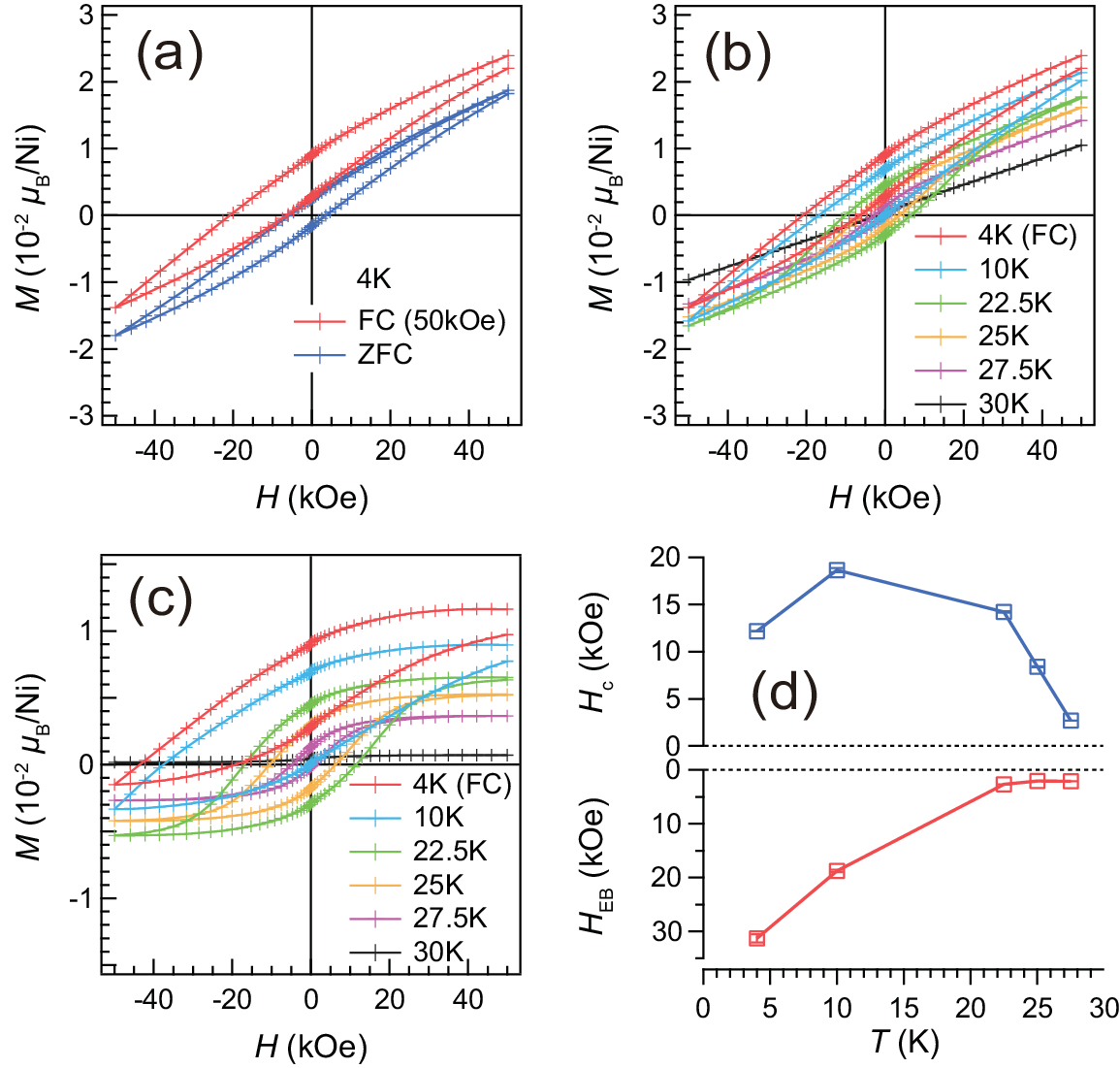}
\end{center}
\caption{(Color online) Magnetic hysteresis and exchange bias in NiS$_2$ NCs. (a) \(M\)-\(H\) curves at 4 K after ZFC and FC processes. (b) Temperature-dependent \(M\)-\(H\) curves measured after field cooling to 4 K. (c) Background-subtracted hysteresis loops. (d) Temperature dependences of \(H_{\rm c}\) and \(H_{\rm EB}\).
}
\label{f4}
\end{figure}

Figure~4 summarizes the magnetic hysteresis of NiS$_2$ NCs from 4 K up to 30 K, just above \(T_{\rm N2}\). Figure~4(a) compares the \(M\)-\(H\) curves measured at 4 K after ZFC and FC processes up to \(H=50\) kOe. Both curves show clear hysteresis, and even the ZFC loop has a large coercive field of several kOe. In addition, the magnetization is not fully reversible within the measured field range. Field cooling makes the loop asymmetric in two ways. First, the loop center is shifted along the field axis, corresponding to exchange bias. Second, the loop is shifted along the magnetization axis, corresponding to a vertical shift. Such horizontal and vertical loop shifts are commonly discussed in exchange-biased ferro-/antiferromagnetic systems: the former reflects unidirectional coupling, whereas the latter reflects pinned uncompensated moments that do not fully reverse within the measured field range \cite{Blachowicz}.

Figure~4(b) shows \(M\)-\(H\) curves measured after field cooling to 4 K and then warming to selected temperatures. The 4 K curve is the same as the FC curve in Fig.~4(a). Upon warming, the hysteresis loop becomes progressively smaller, and the horizontal and vertical shifts are also reduced. At 30 K, just above \(T_{\rm N2}\), the \(M\)-\(H\) curve is dominated by an almost linear response up to \(H=50\) kOe, with only a small nonlinear contribution. Thus, the dominant hysteretic ferromagnetic response in the NCs appears only below \(T_{\rm N2}\), indicating that it is tied to the intrinsic low-temperature antiferromagnetic state of NiS$_2$. 

To quantify the hysteretic component, a high-field linear contribution was subtracted from the \(M\)-\(H\) curves in Fig.~4(b). The background-subtracted loops are shown in Fig.~4(c). Finite remanence and coercivity are observed below \(T_{\rm N2}\), whereas at 30 K the hysteretic component is essentially absent and only a small vertical offset remains. The exchange-bias field \(H_{\rm EB}\) and coercive field \(H_{\rm c}\) were estimated from the positive and negative coercive fields, \(H_{\rm c}^{+}\) and \(H_{\rm c}^{-}\), as
\[
H_{\rm EB}= -\frac{H_{\rm c}^{+}+H_{\rm c}^{-}}{2},
\qquad
H_{\rm c}= \frac{H_{\rm c}^{+}-H_{\rm c}^{-}}{2}.
\]
With this convention, the negative field shift of the FC loop corresponds to a positive \(H_{\rm EB}\).

As shown in Fig.~4(d), both \(H_{\rm EB}\) and \(H_{\rm c}\) appear only below \(T_{\rm N2}\). \(H_{\rm EB}\) decreases more rapidly on warming, whereas \(H_{\rm c}\) remains sizable up to approximately 22.5 K. Such distinct temperature dependences are commonly observed in exchange-biased nanostructures and reflect different thermal stabilities of the unidirectional bias and local reversal barriers  \cite{Blachowicz,Obaidat}. The nonmonotonic low-temperature variation of \(H_{\rm c}\) may reflect a thermally assisted change in the reversal process.

The ZFC peak just below \(T_{\rm N2}\), already visible in Fig.~2, shifts to lower temperature when the magnetic field is increased from 100 Oe to 10 kOe (not shown). This field dependence, together with the absence of an additional sharp heat-capacity anomaly in Fig.~3, indicates that the peak is not a separate thermodynamic phase transition. Because the peak appears in the same temperature range where \(H_{\rm c}\) decreases rapidly in Fig.~4(d), it is likely associated with thermally assisted reversal or depinning of the ferromagnetic component responsible for the hysteresis in the NCs.

Having established that the dominant hysteretic response appears only below \(T_{\rm N2}\), we next consider its implication for the origin of weak ferromagnetism in NiS$_2$. The magnitude of the ferromagnetic response in the NCs is smaller than that reported for high-quality bulk single crystals, but it remains sizable. A rough comparison with the \(M\)-\(H\) curve of bulk NiS$_2$ at 10 K reported by Higo and Nakatsuji suggests a bulk ferromagnetic component of order \(0.02~\mu_{\rm B}/{\rm Ni}\) after subtracting the high-field linear contribution \cite{Higo}. In the present NCs, the corresponding ferromagnetic component is of order \(0.01~\mu_{\rm B}/{\rm Ni}\) when the vertical-shift contribution is included. Thus, the ferromagnetic response is reduced by nanocrystallization, but its remaining magnitude appears too large to be naturally assigned to an internal domain-wall contribution alone, whose formation should be suppressed in nanoscale particles. On the other hand, within a homogeneous-canting picture, a moderate reduction of the ferromagnetic moment could result from local structural modifications, disorder of the antiferromagnetic order near the surface, or a reduction in the canting angle caused by nanocrystallization. We therefore examine whether the observed ferromagnetic response can be understood as the survival of a homogeneous canted moment.

The exchange-bias shift, however, is difficult to reconcile with a simple homogeneous-canting picture. In such a picture, the ferromagnetic moment is a uniform component of the same low-temperature antiferromagnetic order, rather than a separate ferromagnetic component exchange-coupled to a pinned antiferromagnetic region, as required for exchange bias \cite{Blachowicz}. A homogeneous-canting model is therefore not expected to produce an exchange-bias shift by itself. On the other hand, the large \(H_{\rm c}\) and vertical loop shift may, in principle, arise even within a homogeneous-canting picture. Interparticle interactions can enhance \(H_{\rm c}\), although their effect on a volume-distributed moment through limited particle contacts is unclear \cite{Blachowicz,Boekelheide,Gomes}. Moreover, suppression of the \(90^\circ\) domain-wall displacement responsible for reversal in bulk NiS$_2$ may favor quasi-coherent rotation in single-domain NCs \cite{Nagata}. The reported anisotropy-field scale of at least 19--21 kOe is comparable to the maximum \(H_{\rm c}=18.7\) kOe observed here \cite{Nagata,Adachi}, although thermal activation associated with the reduced particle volume acts in the opposite direction.

Another possible origin of ferromagnetic moments in NCs is surface-step-related magnetism. This possibility is relevant because exchange-bias-type effects have recently been reported in NiS$_2$ nanoflakes and attributed to ferromagnetic regions formed near surface step edges and coupled to antiferromagnetic regions \cite{Hartmann}. However, this scenario also cannot account for the dominant hysteretic response observed here. If step-edge moments were the dominant ferromagnetic component, the disappearance of the low-temperature antiferromagnetic order at \(T_{\rm N2}\) could reduce their coercivity or thermally unblock them, but either a hysteretic response with reduced coercivity or a reversible nonlinear \(M\)-\(H\) response should remain above \(T_{\rm N2}\) \cite{Barrera}. However, the observed curve at 30 K is nearly linear, apart from a small vertical offset. Moreover, at \(H=1000\) Oe, the FC--ZFC difference in Fig.~2 increases by approximately \(3.2\times10^{-4}\) emu/mol-Ni/Oe between 70 and 30 K, whereas it increases by approximately \(5.7\times10^{-3}\) emu/mol-Ni/Oe between 30 and 4 K. Together, these results indicate that step-edge moments are not the dominant ferromagnetic component and that the major part of the irreversible magnetic response develops below \(T_{\rm N2}\). The ZFC--FC bifurcation around 70 K may instead reflect a minor frozen surface-step-related component, which may also account for the small vertical shift of approximately \(4.4\times10^{-4}~\mu_{\rm B}/{\rm Ni}\) retained in the background-subtracted curve at 30 K in Fig.~4(c).

The remaining natural explanation is that the dominant ferromagnetic component in the NCs is produced at the particle surface when the low-temperature antiferromagnetic order develops below \(T_{\rm N2}\). Unlike pre-existing step-edge moments, such uncompensated surface moments are induced by the antiferromagnetic order itself and therefore naturally appear only below \(T_{\rm N2}\). Ferromagnetic-like hysteresis from uncompensated surface spins is well known in antiferromagnetic nanoparticles such as NiO, where surface spin uncompensation and surface-spin freezing have been shown to produce sizable moments, coercivity, and exchange-bias-like behavior \cite{Winkler,Shahzad}. Because the moments are located at the particle surface, they can be strongly affected by exchange coupling both to the antiferromagnetic interior of the same NC and to neighboring NCs at particle contacts. This geometry provides a natural route to the large \(H_{\rm c}\), the exchange-bias shift, and the vertical loop shift observed after field cooling. A simple estimate also shows that this interpretation is quantitatively plausible. For a 27 nm particle, one Ni surface layer corresponds to about 6\% of all Ni sites. The ferromagnetic component estimated from the 4 K \(M\)-\(H\) loop is of order \(10^{-2}\,\mu_{\rm B}/{\rm Ni}\). This value can be accounted for by an average uncompensated moment of only \(\sim0.1\)--\(0.2~\mu_{\rm B}\) per surface Ni.

The surface-uncompensated-moment picture indicates that the bulk-like weak-ferromagnetic component is strongly suppressed in the NCs. The background-subtracted loops in Fig.~4(c) do not show clear evidence for multiple independent ferromagnetic components with distinct switching fields; rather, the hysteresis is dominated by a single component. Thus, the uncompensated surface moment accounts for the main ferromagnetic hysteresis in the NCs, while any additional contribution from a bulk-like weak-ferromagnetic component must be minor. This is what is expected if the weak ferromagnetism of bulk NiS$_2$ is carried by extended antiferromagnetic domain walls, because such domain-wall moments should be suppressed when the crystal size is reduced to the nanometer scale. Therefore, the strong suppression of a bulk-like weak-ferromagnetic component indirectly favors the domain-wall scenario for the weak ferromagnetism of bulk NiS$_2$.

In summary, structurally well-ordered NiS$_2$ NCs retain the bulk-like two-step antiferromagnetic transitions, while a pronounced hysteretic ferromagnetic response with large coercivity, exchange bias, and a vertical loop shift develops only below \(T_{\rm N2}\). Taken together, these features are best explained by uncompensated surface moments generated where the low-temperature antiferromagnetic order terminates at the NC surface. The dominance of this surface-derived component and the strong suppression of an additional  bulk-like weak-ferromagnetic component indirectly favors a domain-wall scenario for the weak ferromagnetism of bulk NiS$_2$.

\begin{acknowledgments}
The XRD experiments at SPring-8  were performed with the approval of JASRI (Proposal Nos. 2023B4133, 2025A4140, and 2025B4138). This work was supported by a Japan Society for the Promotion of Science Grant-in-Aid for Scientific Research (C), No. 25K08419 and  by Institute of Industrial Nanomaterials, Kumamoto University. TEM analyses were conducted with a JEM-1400 microscope at the Analytical Research Center for Experimental Sciences, Saga University.

\end{acknowledgments}

\end{document}